\documentclass{ptapap}

\usepackage{amsthm}
\usepackage{amssymb}
\usepackage{amsmath}
\usepackage{color}
\usepackage{mathabx}
\newcommand{\annot}[1]{{\textbf{\textcolor{red}{#1}}}}
\usepackage{soul}

\author{Michał Bejger}[CAMK]

\affil[CAMK]{Nicolaus Copernicus Astronomical Center, Polish Academy of Sciences, Bartycka 18, 00--716 Warszawa, Poland}

\title{Neutron stars as sources of gravitational waves} 

\begin{document}

\maketitle

\begin{abstract}
The global network of ground-based gravitational-wave detectors (the Advanced
LIGO and the Advanced Virgo) is sensitive at the frequency range corresponding
to relativistic stellar-mass compact objects. Among the promising types of
gravitational-wave sources are binary systems and rotating, deformed neutron
stars. I will describe these sources and present predictions of
how their observations will contribute to modern
astrophysics in the near future. 
\end{abstract}

\section{Introduction}
\label{sect:intro} 

Last two years were a turning-point for the gravitational-wave (GW)
astrophysics. First direct detections of GWs from binary systems of massive
stellar black holes with the Advanced LIGO detectors (\citealt{GW150914} and
subsequent ones) demonstrated the viability of the detection principle using
the kilometer-size laser interferometry. This state-of-the-art (Advanced LIGO
\citealt{ALIGO2015} and Advanced Virgo \citealt{AdV2015}) technology creates an
unprecedented opportunity for studying the Universe through a novel 
channel of spacetime distortion \citep{Einstein1916}
measurements. For a mainstream astronomer, GW astronomy may seem a rather
non-standard way of studying the Universe -- it is more like `listening to' than
simply 'looking at' the skies. Motivated by the choice of potential
sources, the ground-based GW detectors of LIGO and Virgo are sensitive to the
range of frequencies similar to the audible range of human ears -- between 10 Hz
(limited by the seismic noise) and a few kHz (limited by the quantum nature of
laser light). As in the case of an ear, a solitary laser interferometer is
practically omnidirectional (has a poor angular resolution), and has no imaging
capabilities; the localisation of sources is performed using a global network
of detectors, currently realized by the three detectors of the LIGO-Virgo
Collaboration.  

GWs are created by a bulk movement of large, rapidly-moving masses: their accelerated
movement provides a time-varying quadrupole, which is the lowest
radiating moment in the theory of general relativity (GR). 
Once emitted, GWs are
weakly coupled to the surrounding matter and propagate freely without
scattering. This has to be contrasted with the electromagnetic emission that
originates at the microscopic level,  is strongly coupled to the surroundings
and often reprocessed; it carries a reliable information from the last
scattering surface only. GWs are therefore the perfect counterparts to the
electromagnetic waves as they provide us with the information impossible to
obtain by other means. 

Science needed 100 years since the birth of GR \citep{Einstein1915} to the
moment when GWs are routinely directly registered by sophisticated apparatus on
the Earth, and changing our understanding of the Universe. The Advanced LIGO
and Advanced Virgo detectors were build thanks by part to the {\it indirect}
evidence gathered in early times of the development of the theory and
observations. In contrary to the Newton's theory of gravity, in which the state
of moving masses does not evolve when no dissipation mechanisms are present, GR
has the energy dissipation mechanism build-in into its very fabric. Early
seminal works of prof. Andrzej Trautman convincingly showed that GWs carry the
energy away from the radiating system, and thus are real physical phenomena and
not artifacts of coordinate choice \citep{RobinsonT1960}. From the
observational side, prof. Bohdan Paczyński demonstrated that extremely short
orbital periods of binaries containing white dwarfs, WZ Sge and HZ29, are
possible only due to the emission of GWs \citep{Paczynski1967}. 
Full realization that binary systems do in fact evolve according to GR came with the
discoveries of binary neutron star systems in the 70s (for a summary see,
e.g., \citealt{WeisbergT2005}). Every binary system emits GWs, leading to the
tightening of its orbit and increasing its orbital frequency, inevitably
driving the components to merge. We may therefore expect that there exist an
astrophysically interesting category of {\it cataclysmic} sources present in
the sensitivity band of detectors only for a limited time: inspiralling and
merging black-hole and neutron-star systems, as well as the core-collapse
supernov{\ae}.
 
Neutron stars are among prime targets for the GW searches. They are less
compact than black holes, but instead of being pure manifestation of the
spacetime curvature they are composed of the most extreme matter existing
currently in the Universe. Thus they provide truly unique conditions to study
matter at the highest densities, pressures, in the presence of the most
powerful magnetic fields and in the regime of strong gravity. These conditions
cannot be reproduced (or even approximated) in the terrestrial laboratories.
Neutron stars are involved in the most spectacular astrophysical phenomena like
supernov{\ae}, or magnetars  and gamma-ray bursts, yet very little is known
about their internal microscopical composition, maximum masses and spins,
radii, and other parameters.

\section{Binary systems} 
\label{sect:cbc} 

I will first focus on the binary systems, since they were the first sources to
be detected. I will use a simplified Newtonian description to describe the basic 
parameters one can infer from the inspiral part of the GW signal. 
 
GW amplitude $h$ (the spacetime distortion, ''the strain'') is proportional to $1/r$,
with $r$ being the luminosity (or rather `loudness') distance to the source.
This relation is a direct consequence of the conservation of energy. For a
binary system of masses $m_1$ and $m_2$ at semi-major axis $a$, with the total
mass $M=m_1 + m_2$, reduced mass $\mu=m_1m_2/M$, and quadrupole moment
$Q\,{\propto}\, \mu a^2$, $h$ is proportional to the second time derivative of $Q$, 
representing the accelerated movement of masses:
$h\,{\propto}\,\ddot{Q}/r\,{\propto}\,\mu a^2 \omega^2/r$. In more detail 
\citep{Einstein1918}, $h \simeq G^{5/3}\mu M^{2/3}\omega^{2/3}/(c^4r)$, where
we make use of Kepler's third law ($GM=a^3\omega^2$). 

Similarly, the GW luminosity $\mathcal{L}$ (the rate of GW-related energy loss,
integrated over a sphere at a distance $r$) is $\mathcal{L} =
dE_{\rm GW}/{dt}\,{\propto}\, G h^2 \omega^2 / c^5\,{\propto}\, G \mu^2 a^4 \omega^6/
c^5$ using the dimensional analysis arguments.  Waves leave the system at the
expense of its orbital energy $E_{\rm orb} = - Gm_1m_2/(2a)$, yielding
$dE_{orb}/{dt} \equiv {Gm_1m_2}\dot{a}/(2a^2) = -{dE_{\rm GW}}/{dt}$. From the time
derivative of the third Kepler's third law, $\dot{a} =
-2a\dot{\omega}/\left(3\omega\right)$, we get the evolution of the orbital
frequency driven by the gravitational-wave emission: $\dot{\omega} =
(96/5)G^{5/3}\omega^{11/3}\mathcal{M}^{5/3}/c^5$.   

The system changes by increasing its orbital frequency; at the same time the
strain amplitude $h$ of emitted waves also grows. This characteristic
frequency-amplitude evolution is called the {\it chirp}, and the characteristic
function of component masses $\mathcal{M} = \left(\mu^3M^2\right)^{1/5} = (m_1
m_2)^{3/5}/(m_1 + m_2)^{1/5}$ is called the {\it chirp mass}. Orbital frequency
is related in a straightforward manner to the GW frequency
$f_{\rm GW}$: from the geometry of the problem it is evident that the frequency of
radiation is predominantly at twice the orbital frequency, $f_{\rm GW} =
\omega/\pi$. The chirp mass $\mathcal{M}$ is therefore directly measured by the
detector registering the evolution of $f_{\rm GW}$; may be e.g., recovered the time-frequency
spectrograms: 

\begin{equation} 
  \mathcal{M} = \frac{c^3}{G}\left(\frac{5}{96}\pi^{-8/3}
  f^{-11/3}_{\rm GW}\dot{f}_{\rm GW}\right)^{3/5}.  
\end{equation}
Time evolution of $\dot{f}_{\rm GW}$ and $h$ give also the distance to the
source $r$. Again, since it is a function of the amplitude and frequency parameters, 
the loudness distance $r$ is a directly measurable quantity: 

\begin{equation} 
  r = \frac{5}{96\pi^2} \frac{c}{h}\frac{\dot{f}_{\rm GW}}{f^3_{\rm GW}}. 
\end{equation}
These properties make the merging binary systems analogues to standard
candles of traditional astronomy, hence they are sometimes called ''standard
sirens'' \citep{HolzH2005}. The idea of using well-understood signals to infer
the distance and constrain the cosmological parameters was proposed first by 
\citet{Schutz1986}. 

The GW signal is extracted from the noisy data using matched-filter techniques,  
correlating the signal model with the data as it evolves in the sensitivity band, 
which means that following the phase of the signal is crucial in the process. For wide 
binary systems, point particles or black holes, the frequency-domain phase $\Psi$ may 
be expanded in the following series of a small parameter: 
\begin{equation} 
  \Psi(f_{\rm GW}) \equiv \Psi_{\rm PP}(f_{\rm GW})\,{\propto}\, \frac{3M}{128\mu v^{5/2}} 
  \sum_{k=0}^N \alpha_{\rm k} v^{k/2},
  \label{eq:phase} 
\end{equation}
that in this example is the orbital velocity $v\,{\propto}\,
\left(\pi\mathcal{M}f_{\rm GW}\right)^{1/3}$.  At cosmological distances, the
observed frequency $f_{GW}$ is redshifted by the expansion of the Universe by
$(1+z)$ (that is, $f_{\rm GW}\to f_{\rm GW}/(1+z)$). 
Purely vacuum GR is not equipped with the
in-build mass scale, so if the source doesn't emit light (like in the case of
black holes), there is no way of breaking the redshift degeneracy between the
$f_{\rm GW}$ and $\mathcal{M}$. 

Fortunately, in case of binary neutron stars which are without a doubt material
sources, the point-particle description breaks down sufficiently early before
the merger, for additional effects related to interactions between the bodies
could be detected. Early part of the inspiral is dominated by gravitational
back reaction (a function of component masses and spins). In the late
inspiral however, tidal effects (mutual deformation of stars in the
gravitational field of the companion) become important. A relation between the
tidal tensor $\mathcal{E}_{ij}$ of one of the components inducing quadrupole
moment $Q_{ij}$ in the other is, in the adiabatic approximation, $Q_{ij} =
\lambda(EOS)\mathcal{E}_{ij}$, where the {\it tidal deformability} $\lambda$ 
is a function of the equation of state (EOS) and the mass of the star, $\lambda = 
(2/3)k_2(EOS)R^5$. The quantity $k_2$ is the Love number \citep{Love1911}, whereas 
$R$ is the radius of the star. From the scaling,
 one can deduce \annot that the deformability 
is a high order effect (5th post-Newtonian order in Eq.~\ref{eq:phase}, $k=10$), so it
becomes large enough only for very tight, relativistic binary systems a few
orbits before the merger. Measurement of the components' $\lambda$s can reveal,
in addition to already measured masses, the radius and compressibility of the
star. Moreover, the fact that the components deform each other during inspiral
changes the behavior of the phase of the GW signal: $\Psi \neq \Psi_{\rm PP}$ for
sufficiently tight systems, and must be complemented by an additional term
related to the tidal deformation, $\Psi_{\rm tidal}$. This additional term is a
function of the Love numbers, masses and radii of stars, hence it breaks the
degeneracy in the expansion in the small parameter $v$ in Eq.~\ref{eq:phase}.
In principle, the measurement of tidal terms allows to measure
the loudness distance and establish the cosmological redshift at the same time.    

\section{Persistent gravitational radiation} 
\label{sect:cw} 

Second category are the {\it continuous} sources, e.g. very wide binary
systems, or rotating, deformed or oscillating neutron stars (''GW pulsars'').
These sources produced long-lived GWs (duration $T$ is much longer than the
observing time $T_{\rm obs}$) and are nearly periodic, with the GW frequency
usually proportional to the characteristic rotational frequency of the object,
$f_{\rm GW}\, {\propto}\, f_{\rm rot}$. In these type of sources, mechanisms 
responsible for creating a time-varying quadrupole are related to deformations
sustained by elastic and/or magnetic stresses (''mountains'', $f_{\rm GW} =
2f_{\rm rot}$), unstable Rosby modes driven by the Coriolis force ($f_{\rm GW} =
4/3f_{\rm rot}$), free precession ($f_{\rm GW}\, {\propto}\, f_{\rm rot} + f_{\rm prec}$) or
accretion contributing to deformation via thermal gradients ($f_{\rm GW}\simeq
f_{\rm rot}$) (for a recent overview of the modeling of periodic GWs see
\citealt{Lasky2015}). GW amplitude (strain) $h_{\rm 0}$ for these sources is
proportional to the degree of asymmetry and the spin frequency,
$h_0\,{\propto}\, I_3\epsilon f^2$, where $\epsilon = (I_1 - I_2)/I_3$ is the
deformation, and $I_i$ are the value of the moment of inertia along the
principal axes, $I_3$ being aligned with the rotation axis.  Depending on the
dense-matter models, $\epsilon_{\rm max} = 10^{-3} - 10^{-6}$.  Recent results on
the LIGO O1 data \cite{Abbott-allskyO1,Abbott-known-pulsarsO1} did not provide
direct detections (yet), but physically interesting upper limits were obtained
for several known sources and across a wide frequency range. 
The upper limits for GW radiation are compared with
the available reservoir of energy, rotational energy $E_{\rm rot}$. Rotational
energy loss $\dot{E}_{rot}\,{\propto}\, f\dot{f}$, and the energy emitted in
GWs scales like $\dot{E}_{\rm GW}\,{\propto}\, f^6 I_3^2\epsilon^2$.  Assuming
$\dot{E}_{\rm rot} = \dot{E}_{\rm GW}$, the knowledge of $I_3$ and the distance to
the source, we get the spin-down upper limit for the strain,
$h_{\rm 0}^{\rm sd}\,{\propto}\, \sqrt{\lvert \dot{f}\rvert I_3/f}$. Recent results show
that the Crab pulsar emits less than $2\times 10^{-3}\ \dot{E}_{\rm rot}$ in GWs,
and the Vela pulsar less than $10^{-2}\ \dot{E}_{\rm rot}$. These upper limits
improve previously obtained by a factor of 2.5 and 
allow for excluding extreme radiation models (for more details on the status
of the data analysis of continuous GWs from rotating neutron stars see
\citealt{Bejger2017}). 

Persistent GWs are also expected in the form of stochastic background produced
by the populations of (continuous or transient) sources, or even by GWs created
in the very early Universe. 

\section{The dawn of multi-messenger astronomy} 

GWs and photons provide complementary information about the physics of the
source and its environment. From GWs we may learn about the masses, spins and
eccentricity in case of binary systems, and deformation in case of rotating
objects, system orientation, distance, and also the size of the population and
the rate at which astrophysical phenomena happen. Electromagnetic observations
provide precise sky localisation, redshifts of the host galaxies, information about the local environment, emission processes and acceleration mechanisms.
Global network of three interferometers of LIGO and Virgo is currently able to
provide sky localisation within $\sim 10$ square degrees from the GW
observations alone (triangulation) which allow for rapid electromagnetic follow-up. At its nominal
sensitivity, the LIGO-Virgo collaboration will be able to routinely detect binary black
hole mergers up to 1 Gpc distance, and neutron-star mergers up to 200 Mpc.  

\section{Afterword: GW170817} 
\label{sec:afterword} 

Judging from the lively discussion during and after my talk, the audience saw
through my slides and was already aware of the spreading rumor of a new
breakthrough discovery, that became public approximately one month later, on
October 16, 2017.  Indeed, August 2017 was an extremely interesting experience:
first binary black-hole merger observation with the global network of three
detectors of Advanced LIGO and the Advanced Virgo \citep{GW170814}, as well as
the first Advanced LIGO and Advanced Virgo detection of a nearby (at the distance
of 40 Mpc) binary neutron-star merger \citep{GW170817}, followed by a short
gamma-ray burst \citep{GW170817GRB} and broad-band electromagnetic emission
observational campaign \citep{GW170817MMA} and the kilonova study. The
detection of a beautiful, surprisingly strong chirp signal from the closest
short gamma-ray burst to date, served as the best proof for theoretical ideas
put forward by prof. Paczyński: connection between short gamma-ray bursts and
neutron-star mergers \citep{Paczynski1986} and the physics of kilonova
\citep{LiP1998}. GW170817 was the best demonstration of techniques described in
this talk: precise triangulation with three GW detectors crucial for
the kilonova follow-up, direct ''standard siren'' (i.e., bypassing traditional 
''distance ladders'') measurement of distance to the host galaxy and
independent measurement of the Hubble constant \citep{GW170817H0}, measurement
of the speed of gravitational waves, and tidal deformabilities of component
neutron stars. Thanks to neutron stars we now witness a true beginning of the GW astronomy.  

\acknowledgements{This work was partially supported by the Polish NCN grants no.
2014/14/M/ST9/00707 and 2016/22/E/ST9/00037.} 

\bibliographystyle{ptapap}
\bibliography{bejger}

\begin{thebibliography}{23}
\providecommand{\natexlab}[1]{#1}
\providecommand{\url}[1]{\texttt{#1}}
\providecommand{\urlprefix}{URL }
\providecommand{\eprint}[2][]{\url{#2}}

\bibitem[{{Aasi} et~al.(2015)}]{ALIGO2015}
{Aasi}, J., et~al., \emph{{Advanced LIGO}}, \emph{\cqg} \textbf{32}, 7, 074001
  (2015), \eprint{1411.4547}

\bibitem[{{Abbott} et~al.(2016)}]{GW150914}
{Abbott}, B.~P., et~al., \emph{{Observation of Gravitational Waves from a
  Binary Black Hole Merger}}, \emph{\prl} \textbf{116}, 6, 061102 (2016),
  \eprint{1602.03837}

\bibitem[{{Abbott} et~al.(2017{\natexlab{a}})}]{GW170817H0}
{Abbott}, B.~P., et~al., \emph{{A gravitational-wave standard siren measurement
  of the Hubble constant}}, \emph{Nature}  (2017{\natexlab{a}}),
  \eprint{1710.05835}

\bibitem[{{Abbott} et~al.(2017{\natexlab{b}})}]{Abbott-allskyO1}
{Abbott}, B.~P., et~al., \emph{{All-sky search for periodic gravitational waves
  in the O1 LIGO data}}, \emph{\prd} \textbf{96}, 6, 062002
  (2017{\natexlab{b}}), \eprint{1707.02667}

\bibitem[{{Abbott} et~al.(2017{\natexlab{c}})}]{Abbott-known-pulsarsO1}
{Abbott}, B.~P., et~al., \emph{{First Search for Gravitational Waves from Known
  Pulsars with Advanced LIGO}}, \emph{\apj} \textbf{839}, 12
  (2017{\natexlab{c}}), \eprint{1701.07709}

\bibitem[{{Abbott} et~al.(2017{\natexlab{d}})}]{GW170817GRB}
{Abbott}, B.~P., et~al., \emph{Gravitational Waves and Gamma-Rays from a Binary
  Neutron Star Merger: GW170817 and GRB 170817A}, \emph{The Astrophysical
  Journal Letters} \textbf{848}, 2, L13 (2017{\natexlab{d}})

\bibitem[{{Abbott} et~al.(2017{\natexlab{e}})}]{GW170814}
{Abbott}, B.~P., et~al., \emph{GW170814: A Three-Detector Observation of
  Gravitational Waves from a Binary Black Hole Coalescence}, \emph{Phys. Rev.
  Lett.} \textbf{119}, 141101 (2017{\natexlab{e}})

\bibitem[{{Abbott} et~al.(2017{\natexlab{f}})}]{GW170817}
{Abbott}, B.~P., et~al., \emph{GW170817: Observation of Gravitational Waves
  from a Binary Neutron Star Inspiral}, \emph{Phys. Rev. Lett.} \textbf{119},
  161101 (2017{\natexlab{f}})

\bibitem[{{Abbott} et~al.(2017{\natexlab{g}})}]{GW170817MMA}
{Abbott}, B.~P., et~al., \emph{Multi-messenger Observations of a Binary Neutron
  Star Merger}, \emph{The Astrophysical Journal Letters} \textbf{848}, 2, L12
  (2017{\natexlab{g}})

\bibitem[{{Acernese} et~al.(2015)}]{AdV2015}
{Acernese}, F., et~al., \emph{{Advanced Virgo: a second-generation
  interferometric gravitational wave detector}}, \emph{\cqg} \textbf{32}, 2,
  024001 (2015), \eprint{1408.3978}

\bibitem[{{Bejger}(2017)}]{Bejger2017}
{Bejger}, M., \emph{{Status of the continuous gravitational wave searches in
  the Advanced Detector Era}}, \emph{Rencontres de Moriond}  (2017),
  \eprint{arXiv:1710.06607}

\bibitem[{{Einstein}(1915)}]{Einstein1915}
{Einstein}, A., \emph{{Zur allgemeinen Relativit{\"a}tstheorie}},
  \emph{Sitzungsberichte der K{\"o}niglich Preu{\ss}ischen Akademie der
  Wissenschaften (Berlin), Seite 778-786.}  (1915)

\bibitem[{{Einstein}(1916)}]{Einstein1916}
{Einstein}, A., \emph{{N{\"a}herungsweise Integration der Feldgleichungen der
  Gravitation}}, \emph{Sitzungsberichte der K{\"o}niglich Preu{\ss}ischen
  Akademie der Wissenschaften (Berlin), Seite 688-696.}  (1916)

\bibitem[{{Einstein}(1918)}]{Einstein1918}
{Einstein}, A., \emph{{{\"U}ber Gravitationswellen}}, \emph{Sitzungsberichte
  der K{\"o}niglich Preu{\ss}ischen Akademie der Wissenschaften (Berlin), Seite
  154-167.}  (1918)

\bibitem[{{Holz} \& {Hughes}(2005)}]{HolzH2005}
{Holz}, D.~E., {Hughes}, S.~A., \emph{{Using Gravitational-Wave Standard
  Sirens}}, \emph{\apj} \textbf{629}, 15 (2005), \eprint{astro-ph/0504616}

\bibitem[{Lasky(2015)}]{Lasky2015}
Lasky, P.~D., \emph{Gravitational Waves from Neutron Stars: A Review},
  \emph{Publications of the Astronomical Society of Australia} \textbf{32}
  (2015)

\bibitem[{{Li} \& {Paczy{\'n}ski}(1998)}]{LiP1998}
{Li}, L.-X., {Paczy{\'n}ski}, B., \emph{{Transient Events from Neutron Star
  Mergers}}, \emph{\apjl} \textbf{507}, L59 (1998), \eprint{astro-ph/9807272}

\bibitem[{{Love}(1911)}]{Love1911}
{Love}, A.~E.~H., {Some Problems of Geodynamics} (1911)

\bibitem[{{Paczy{\'n}ski}(1967)}]{Paczynski1967}
{Paczy{\'n}ski}, B., \emph{{Gravitational Waves and the Evolution of Close
  Binaries}}, \emph{\actaa} \textbf{17}, 287 (1967)

\bibitem[{{Paczynski}(1986)}]{Paczynski1986}
{Paczynski}, B., \emph{{Gamma-ray bursters at cosmological distances}},
  \emph{\apjl} \textbf{308}, L43 (1986)

\bibitem[{{Robinson} \& {Trautman}(1960)}]{RobinsonT1960}
{Robinson}, I., {Trautman}, A., \emph{{Spherical Gravitational Waves}},
  \emph{Physical Review Letters} \textbf{4}, 431 (1960)

\bibitem[{{Schutz}(1986)}]{Schutz1986}
{Schutz}, B.~F., \emph{{Determining the Hubble constant from gravitational wave
  observations}}, \emph{\nat} \textbf{323}, 310 (1986)

\bibitem[{{Weisberg} \& {Taylor}(2005)}]{WeisbergT2005}
{Weisberg}, J.~M., {Taylor}, J.~H., \emph{{The Relativistic Binary Pulsar
  B1913+16: Thirty Years of Observations and Analysis}}, in F.~A. {Rasio},
  I.~H. {Stairs} (eds.) Binary Radio Pulsars, \emph{Astronomical Society of the
  Pacific Conference Series}, volume 328, 25 (2005), \eprint{astro-ph/0407149}

\end{thebibliography}

\end{document}